\documentclass[11pt,a4paper]{article}
\usepackage{latexsym}

\newcommand\meet\wedge

\newcommand{\reals}{{\hbox{\sf I\kern-.14em\hbox{R}}}}

\newcommand{\norm}[1]{\left\|\,#1\,\right\|}
\newcommand{\trnorm}[1]{\norm{#1}_{\rm 1}}

\newcommand{\eqdef}{\stackrel{\rm def}{=}}

\newcommand{\trn}[1]{\trnorm{#1}}

\newcommand{\E}{\mathbf{{E}}}
\newcommand{\F}{\mathcal{F}}

\newtheorem{theor}{Theorem}
\newtheorem{lemm}{Lemma}
\newtheorem{coro}{Corollary}
\newtheorem{fact}{Fact}
\newtheorem{defi}{Definition}
\newcommand{\Com}{{\mathchoice {\setbox0=\hbox{$\displaystyle\rm C$}\hbox{\hbox
to0pt{\kern0.4\wd0\vrule height0.9\ht0\hss}\box0}}
{\setbox0=\hbox{$\textstyle\rm C$}\hbox{\hbox
to0pt{\kern0.4\wd0\vrule height0.9\ht0\hss}\box0}}
{\setbox0=\hbox{$\scriptstyle\rm C$}\hbox{\hbox
to0pt{\kern0.4\wd0\vrule height0.9\ht0\hss}\box0}}
{\setbox0=\hbox{$\scriptscriptstyle\rm C$}\hbox{\hbox
to0pt{\kern0.4\wd0\vrule height0.9\ht0\hss}\box0}}}}
\newcommand{\qed}{\hfill$\Box$}
\begin{document}
\date{} \title{On Rounds in Quantum Communication}
\author{Hartmut Klauck\\
  CWI\\
  P.O. Box 94079\\
  1090 GB Amsterdam, the Netherlands\\
  \tt klauck@cwi.nl} \maketitle

\begin{abstract}
  We investigate the power of interaction in two player quantum
  communication protocols. Our main result is a rounds-communication
  hierarchy for the pointer jumping function $f_k$. We show that $f_k$
  needs quantum communication $\Omega(n)$ if Bob starts the
  communication and the number of rounds is limited to $k$ (for any
  constant $k$).
  Trivially, if Alice starts, $O(k\log n)$ communication in $k$ rounds
  suffices.  The lower bound employs a result relating the relative
  von Neumann entropy between density matrices to their trace distance
  and uses a new measure of information.
  
  We also describe a classical probabilistic $k$ round protocol with
  communication $O(n/k\cdot(\log^{(k/2)}n+\log k)+k\log n)$ in which
  Bob starts the communication.
  
  Furthermore as a consequence of the lower bound for pointer jumping
  we show that any $k$ round quantum protocol for the disjointness
  problem needs communication $\Omega(n^{1/k})$ for $k=O(1)$.
\end{abstract}

\section{Introduction}
Quantum mechanical computing and communication has been studied
extensively during the last decade.  Communication has to be a
physical process, so an investigation of the properties of physically
allowed communication is desirable, and the fundamental
theory of physics available to us is quantum mechanics.

The theory of communication complexity deals
with the question how efficient communication
problems can be solved and has various
applications to lower bound proofs
(introduction to (classical) communication
complexity can be found in \cite{KN97}).
The communication complexity approach to lower bounds consists of
reducing a lower bound proof for some computational model to a
communication complexity lower bound, where several techniques for
such proofs are available, see \cite{KN97} for many examples.

In a quantum protocol (as defined in \cite{Y93}) two players
Alice and Bob each receive an input, and have
to compute some function defined on the pair of
inputs cooperatively. To this end they exchange
messages consisting of qubits, until the result
can be produced by some measurement done by one
of the players (for overviews on quantum communication complexity
see \cite{T99} and \cite{Kl00b}).

A slightly different scenario
proposed in \cite{CB97} and \cite{CDNT97} allows the players to start
the protocol possessing some (input independent) qubits that are
entangled with those of the other player. Due to the ``superdense
coding'' technique of \cite{BW92} in this model 2 classical bits can
be communicated by transmitting one qubit (and using up one EPR pair).
See \cite{Y93,Kl00} for some examples of lower bounds via
communication complexity in the quantum setting.
Unfortunately so far only few ``applicable'' lower bound methods for quantum
protocols are known: the rank lower bound is known to hold for exact
(i.e., errorless) quantum communication \cite{BCW98,BW01}, the
(usually weak) discrepancy lower bound for bounded error protocols \cite{K95}.

One breakthrough result of the field of quantum computing is
Grover's search algorithm that retrieves an item from an unordered
list within $O(\sqrt{n})$ questions \cite{G96},
outperforming every classical algorithm for the
problem. By an application of this search algorithm to communication complexity
in  \cite{BCW98} an upper bound of
$O(\sqrt{n}\log n)$ is shown for the bounded
error quantum communication complexity of the
disjointness problem $DISJ_n$ (both players
receive an incidence vector of a subset of
$\{1,\ldots,n\}$ and have to decide whether the
sets are not disjoint: $\bigvee (x_i\wedge
y_i)$), one of the most important communication
problems. This yields the largest gap between
quantum and classical communication complexity
known so far for a total function, the
probabilistic communication complexity of
disjointness is $\Omega(n)$ \cite{KS92}.
Currently no superlogarithmic lower bound on the bounded error quantum
communication complexity of the disjointness problem is known.

Unfortunately the protocol for disjointness using Grover search takes $\Theta(\sqrt{n})$
rounds, an unbounded increase in interaction compared to the trivial
protocol communicating $n$ bits in 1 round. Similar phenomena
show up in the polynomial gaps between quantum Las Vegas and probabilistic bounded error
communication complexity for total functions, see
\cite{BCWZ99,Kl00}.\footnote{Exponential gaps between quantum communication complexity and
classical probabilistic communication complexity
are known only for partial functions, and are possible without
interaction  \cite{R99,BCW98}.}

We are interested in the question how
efficient total communication problems can be solved
in the quantum model when the number of rounds
is restricted.  The most severe restriction is
one-way communication, where only a monologue
is transmitted from one player to the other,
who then decides the function value. This has
been investigated e.g.~in \cite{Kl00}, where a lower
bound method based on the so called
VC-dimension is proved, which allows to prove
an exponential advantage for 2 round classical
compared to 1 round quantum communication
complexity.  Kremer \cite{K95} exhibits a complete problem for the
class of problems with polylogarithmic quantum one-way communication
complexity (in the case of bounded error).

In a series of papers (see \cite{DGS87},
\cite{HR93}, \cite{NW93}, \cite{PRV99},
\cite{Kl98}) more general round hierarchies of
the following form are given for classical
protocols: A function $f_k$ (usually the so
called pointer jumping function) has $k$ round
communication complexity $k\log n$ if Alice
starts the communication, but has a much larger
$k$ round communication complexity when Bob
starts. 

Our main result is that $k$-round quantum protocols
need communication $n/2^{2^{O(k)}}-k\log n$ to
compute the pointer jumping function when Bob
starts. So changing the starting player (or reducing the number of
rounds by 1) may result in drastically increased communication also in
the quantum case.

Nayak et.~al.~\cite{NTZ00} have proved a
lower bound of $\Omega(n^{1/k})$ for the quantum communication
complexity of a certain subproblem of pointer jumping, in the
situation when B starts and $k$ rounds are allowed.

We begin our consideration of the complexity of pointer jumping (in
section 5) with the description of a classical randomized
protocol for pointer jumping using communication
$O(n/k\cdot(\log^{(k/2)}n+\log k)+k\log n)$ in
the situation where Bob starts and only $k$
rounds are allowed. This upper bound is close to the known lower
bounds for classical protocols \cite{NW93,Kl00}.

The general strategy of our new lower bound for pointer jumping is to
bound the value of a certain measure of information between the qubits
of one player and the ``next'' pointer in terms of the
analogous quantity for the previous pointer plus the average information on
pointers in possession of the other player.

The mentioned protocol makes clear why the usual measure of
information does not work in this approach. So (after defining the main notions of quantum
computing in section 2 and the model of communication complexity in
section 3) we introduce a new measure of quantum information in section
4. This measure is tied to the usual, von Neumann
measure of quantum information by a theorem, which connects the
trace distance between density matrices to
their relative von Neumann entropy.

The lower bound on pointer jumping implies via
reductions lower bounds for the $k$ round
bounded error quantum communication complexity
of the disjointness problem of the order
$\Omega(n^{1/k})$ for all constant $k$, see section 6.

We conclude from our result that quantum
communication is dependent on interaction, as one
should expect for a ``realistic'' mode of
communication. We also conclude that good speedups by quantum
protocols imply the use of
nontrivial interaction in the case of total
functions: for an asymptotic speedup by
quantum Las Vegas protocols always more than
one round is necessary \cite{Kl00}. By the
results in this paper (and similar results in \cite{NTZ00})
rounds are also crucial in quantum speedups
for the disjointness problem.

The lower bound for a subproblem of pointer jumping given in ~\cite{NTZ00} and the
 lower bound in this paper use at the basis of the proofs similar
techniques. The main ingredient of the proof in
\cite{NTZ00}, the ``average encoding
theorem'', follows directly from our theorem 1, which
states a connection between a new measure of
information (based on the trace distance) and
the relative von Neumann entropy. We make use of
a fact from \cite{NTZ00}, namely the ``local transition theorem''.
A combined version of both papers appears in \cite{KNTZ01}.

Our results also hold in the model, in which prior entanglement is
available.

The paper is organized as follows: In the next section we give some
background on quantum mechanics. Then we define the communication
model in section 3. In section 4 we consider measures of information
and entropy. In section 5 we prove our results on the complexity of
pointer jumping. Section 6 contains the lower bound for the
disjointness problem.

\section{Quantum States and Transformations}
Quantum mechanics is a theory of reality in terms of states and
transformations of states. See \cite{NC00,Pr} for general information
on this topic with an orientation on quantum computing.

In quantum mechanics pure states are unit norm vectors in a Hilbert
space, usually $\Com^k$. We use the Dirac notation for pure states. So
a pure state is denoted $|\phi\rangle$ or
$\sum_{x\in\{0,\ldots,k-1\}}\alpha_x|x\rangle$ with
$\sum_{x\in\{0,\ldots,k-1\}}|\alpha_x|^2=1$ and with
$\{\,|x\rangle\,|x\in\{0,\ldots,k-1\}\}$ being an orthonormal basis of
$\Com^k$.

Inner products in the Hilbert space are denoted
$\langle\phi|\psi\rangle$, outer (matrix valued) products $|\phi\rangle\langle \psi|$.

If $k=2^l$ then the basis is also denoted
$\{\,|x\rangle\,|x\in\{0,1\}^l\}$. In this case the space $\Com^{2^l}$
is the $l$-wise tensor product of the space $\Com^2$. The latter space
is called a qubit, the former space consists of $l$ qubits.

Usually also mixed states are considered.

\begin{defi}
  Let $\{(p_i,|\phi_i\rangle)|i=1,\ldots,k\}$ with $\sum_i p_i=1$ and
  $p_i\in[0,1]$ be an ensemble of pure states of a quantum system,
  also called a mixed state.
  
  $\rho_i=|\phi_i\rangle\langle\phi_i|$ is the density matrix of the
  pure state $|\phi_i\rangle$. $\sum_i p_i\rho_i$ is the density
  matrix of the mixed state.
  
  For a bipartite system with density matrix $\rho_{AB}$ denote
  $\rho_A=trace_B\rho_{AB}$.
\end{defi}

As usual measurements of certain observables and unitary
transformations are considered as basic operations on states, see
\cite{NC00, Pr} for definitions.

For all possible measurements on a mixed state the results are
determined by its density matrix. In quantum mechanics the density
matrix plays an analogous role to the density function of a random
variable in probability theory.  Note that a density matrix is
Hermitian, positive semidefinite and has trace 1. Thus it has only
real, nonnegative eigenvalues that sum to 1.

Linear transformations on density matrices are called superoperators.

\begin{defi}
  A superoperator is a linear map on density matrices. A superoperator
  is positive, if it sends positive semidefinite matrices to positive
  semidefinite matrices. A superoperator is called completely
  positive, if its tensor product with the identity superoperator is
  positive on density matrices over each finite dimensional extension
  of the underlying Hilbert space.
\end{defi}

Trace-preserving completely positive superoperators map density
matrices to density matrices and capture all physically allowed
transformations. These include unitary transformations, tracing out
subsystems, forming a tensor product with some constant qubits, and
general measurements.

The following important fact characterizes the allowed superoperators
in terms of unitary transformations and tracing out (see \cite{Pr}).
This fact is known as the Kraus representation theorem.

\begin{fact}
  The following statements are equivalent:
\begin{enumerate}
\item A superoperator $T$ sending density matrices over $H_1$ to
  density matrices over $H_2$ is trace preserving and completely
  positive.
\item There is a Hilbert space $H_3$ with $dim(H_3)\le dim(H_1)$ and a
  unitary transformation $U$, such that for all density matrices
  $\rho$ over $H_1$ the following holds:
\[T\rho= trace_{H_1\otimes H_3}
[U(\rho\otimes|0_{H_3\otimes H_2}\rangle\langle0_{H_3\otimes H_2}|)U^\dagger ].\]
\end{enumerate}
\end{fact}

So allowed superoperators can be simulated by adding some blank
qubits, applying a unitary transformation and tracing out, i.e.,
``dropping some qubits''.

\begin{defi}
  A {\it purification} of a mixed state with density matrix $\rho$
  over some Hilbert space $H$ is any pure state $|\phi\rangle$ over
  some space $H\otimes K$ such that $trace_K
  |\phi\rangle\langle\phi|=\rho$.
\end{defi}

\section{The Communication Model}

In this section we provide definitions of the computational models
considered in the paper.  We begin with the model of classical
communication complexity.

\begin{defi}
  Let $f:\{0,1\}^n\times\{0,1\}^n\to\{0,1\}$ be a function. In a
  communication protocol player Alice and Bob receive $x$ and $y$ and
  compute $f(x,y)$.  The players exchange binary encoded messages. The
  communication complexity of a protocol is the worst case number of
  bits exchanged for any input. The deterministic communication
  complexity $D(f)$ of $f$ is the complexity of an optimal protocol
  for $f$.
  
  In a randomized protocol both players have access to public random
  bits. The output is required to be correct with probability
  $1-\epsilon$ for some constant $\epsilon$. The randomized
  communication complexity of a function $R_\epsilon(f)$ is then
  defined analogously to the deterministic communication complexity.
  We define $R(f)=R_{1/3}(f)$.
  
  A protocol has $k$ rounds, if the players exchange $k$ messages with
  Alice and Bob alternating as speakers. In message $k+1$ one player
  announces the result. Alice protocol is called one-way if only one
  players sends a message, and then the other player announces the
  result. The complexity notations are superscripted with the number
  of allowed rounds and eventually with the player starting, like
  $D^{k,B}$ or $D^k$ (usually Alice starts).
\end{defi}

Now we define quantum communication protocols. For general information
on quantum computation see \cite{NC00} and \cite{Pr}.

\begin{defi}
  In a quantum protocol both players have a private set of qubits.
  Some of the qubits are initialized to the input before the start of
  the protocol, the other qubits are in state $|0\rangle$. In a
  communication round one of the players performs some unitary
  transformation on the qubits in his possession and then sends some
  of his qubits to the other player (the latter step does not change
  the global state but rather the possession of individual qubits).
  The choice of qubits to be sent and of unitary operations is fixed
  in advance by the protocol.
  
  At the end of the protocol the state of some qubits belonging to one
  player is measured and the result is taken as the output. The
  communication complexity of the protocol is the number of qubits
  exchanged.
  
  In a (bounded error) quantum protocol the correct answer must be
  given with probability $1-\epsilon$ for some $1/2>\epsilon>0$. The
  (bounded error) quantum complexity of a function, called
  $Q_\epsilon(f)$, is the complexity of an optimal protocol for $f$.
  $Q(f)=Q_{1/3}(f)$.
\end{defi}

In fact we will consider a more general model of communication
complexity, in which the players can apply all physically allowed
superoperators to their private qubits. But due to the Kraus
representation theorem (see fact 1) this model can be simulated by the
above model without increasing communication (with the help of
additional private qubits).

We have to note that in the defined model no intermediate measurements
are allowed to control the choice of qubits to be sent or the time of
the final measurement. Thus for all inputs the same amount of
communication and rounds is used.  As a generalization one could allow
intermediate measurements, whose results could be used to choose the
qubits to be sent and possibly when to stop the communication
protocol. One would have to make sure that the receiving player knows
when a message ends.

A protocol with $k$ rounds in this more general model can be simulated
while loosing a factor of at most $k$ in the communication: for each
measurement the operations given by the Kraus representation theorem
are used. The measurement's result is then stored in some ancilla
qubits. Now the global state is a superposition over the results and a
superposition of the appropriate communications can be used as a
communication. This superposition uses as many qubits as the worst
case message of that round. This may be at most the complexity of the
whole protocol, so the overall complexity increases by at most a
factor of $k$. While this simulation may not be satisfactory in
general, it suffices to keep our lower bound valid in the more general
model.

In \cite{CB97} and \cite{CDNT97} a different model of quantum
communication (the communication model with entanglement) is proposed.
Alice and Bob may possess an arbitrary input-independent set of
(entangled) qubits in the beginning. Then they communicate according
to an ordinary quantum protocol.  This model can be simulated by
allowing first an arbitrary input-independent communication with no
cost followed by a usual quantum communication protocol in which the
cost is measured. The superdense coding technique of \cite{BW92}
allows to transmit $n$ bits of classical information with $\lceil
n/2\rceil$ qubits in this model.

\begin{defi} The quantum bounded error communication complexity with
entanglement and error $\epsilon$ is denoted $Q^{pub}_\epsilon(f)$.
Let $Q^{pub}(f)=Q^{pub}_{1/3}(f)$.
\end{defi}

For surveys on quantum communication complexity see \cite{T99} and
\cite{Kl00b}.

\section{Quantum Information Theory}

Our main result in the next section uses information theory arguments.
First we define the classical notions of entropy and information.

\begin{defi}
  Let $X:\Omega\to S$ be a random variable on finite sets $\Omega, S$
  (as usual the argument of $X$ is dropped). The density function (or
  distribution) of $X$ is $p_X:S\to[0,1]$, where $p_X(x)$ is the
  probability of the event $X=x$.
  
  The entropy of $X$ is $H(X)=-\sum_{x\in S} p_X(x)\log p_X(x)$.
  
  Let $X,Y$ be random variables over $\Omega$. The joint density
  function of $XY$ is $p_{XY}(x,y)$. The information between $X$ and
  $Y$ is $H(X:Y)=H(X)+H(Y)-H(XY)$.
  
  We use the convention $0\log 0=0$.
\end{defi}
Now we define the quantum mechanical notions of entropy and
information.

\begin{defi}
  The von Neumann entropy of a density matrix $\rho_X$ is
  defined by $S(X)=S(\rho_X)=-trace(\rho_X\log\rho_X)$. The relative
  von Neumann entropy between two density matrices $\rho,\sigma$ of
  the same size is $S(\rho||\sigma)=trace(\rho(\log\rho-\log\sigma))$.
  This value may be infinite.
  
  The von Neumann information is $S(X:Y)=S(X)+S(Y)-S(XY)$ (see also
  \cite{CA96}). Here $S(X)$ is the von Neumann entropy of the reduced
  density matrix $\rho_X=trace_Y\rho_{XY}$.
  
  The conditional von Neumann information is
  $S(X:Y|Z)=S(XZ)+S(YZ)-S(Z)-S(XYZ)$.
\end{defi}

Note that the von Neumann entropy depends only on the eigenvalues of a
matrix and is thus invariant under unitary transformations.  If the
underlying Hilbert space has dimension $d$ then
the von Neumann entropy of the density matrix is bounded by $\log d$.

Not all relations in classical information theory hold for von Neumann
entropy. The following fact contains the so-called Araki-Lieb
inequality (*) and its consequences, which describes a notable
difference to classical entropy (see \cite{Pr,CA96}).

\begin{fact} For all bipartite states $\rho_{XY}$:
  
  $S(X)+S(Y)\ge S(XY)\stackrel{*}{\ge} |S(X)-S(Y)|$,
  
  $S(X:Y)\le 2S(X)$.
\end{fact}

Then also $S(X:Y|Z)\le S(X:YZ)\le 2S(X)$ holds.

The following is an important property of the von Neumann entropy, see
\cite{Pr}. This property is known as the Lindblad-Uhlmann monotonicity
of the von Neumann entropy.

\begin{fact}
  For all trace-preserving, completely positive superoperators $F$ and
  all density matrixes $\rho,\sigma$:
\[S(\rho||\sigma)\ge S(F\rho||F\sigma).\]
\end{fact}

We are going to introduce another measure of information based on the
distinguishability between a bipartite state and the state described
by the tensor product of its two reduced density matrices. Now we
consider measures of distinguishability. One such measure is the
relative entropy.  For probability distributions the total variational
distance is another useful measure.

\begin{defi} If $p,q$ are probability distributions on $\{1,\ldots,n\}$, then their distance is defined\[||p-q||=\sum_{i=1}^n |p(i)-q(i)|.\]
\end{defi}

The following norm on linear operators is considered in \cite{AKN98}.

\begin{defi}
  Let $\rho$ be the matrix of a linear operator.  Then the trace norm
  of $\rho$, denoted $||\rho||_{\rm 1}$, is the sum of the absolute values
  of the elements of the multiset of all eigenvalues of $\rho$. In
  particular $||\rho||_{\rm 1}=Tr(\sqrt{\rho^\dag\cdot\rho})$.
\end{defi}

Note that the distance $||\rho-\sigma||_{\rm 1}$ is a real value for
Hermitian matrices $\rho,\sigma$. The trace norm has a close relation
to the measurable distance between states as shown in \cite{AKN98}.

\begin{fact}
  For an observable $O$ and a density matrix $\rho$ denote $p_\rho^O$
  the distribution on the outcomes of a measurement as induced by $O$
  on the state $\rho$.

\[||\rho-\sigma||_{\rm 1}=\max_O\{||p_\rho^O-p_\sigma^O||\}.\]
\end{fact}

So two density matrices that are close in the trace distance cannot be
distinguished well by any measurement.

The next lemma is related to fact 4 and follows from fact 1.

\begin{lemm}
  For each Hermitian matrix $\rho$ and each trace-preserving
  completely positive superoperator $F$:

\[||\rho||_{\rm 1}\ge ||F(\rho)||_{\rm 1}.\]
\end{lemm}

We employ the following theorem to bound the trace distance in terms
of relative entropy. A classical analogue of the theorem can be found
in \cite{Bl87} and has been used e.g.~in \cite{R98}.

\begin{theor}
  For density matrices $\rho,\sigma$ of the same size:

\[S(\rho||\sigma)\ge \frac{1}{2\ln 2} ||\rho-\sigma||_{\rm 1}^2.\]
\end{theor}

{\sc Proof:} Since both the norm and the relative entropy are
invariant under unitary transformations we assume that the basis of
the density matrices diagonalizes $\rho-\sigma$. Note that in general
neither $\rho$ nor $\sigma$ are diagonal now.  Let $S$ be the multiset
of all nonnegative eigenvalues of $\rho-\sigma$ and $R$ the multiset
of all its negative eigenvalues. All eigenvalues are real since
$\rho-\sigma$ is Hermitian. Now if the dimension of the space $H_S$
spanned by the eigenvectors belonging to $S$ has dimensions $k$ and
the space $H_R$ spanned by the eigenvectors belonging to $R$ has
dimensions $n-k$, then increase the size of the underlying Hilbert
space so that both spaces have the same dimension $n'=\max\{k,n-k\}$.
The density matrices have zero entries at the corresponding positions.
Now we view the density matrices as density matrices over a product
space $H_2\otimes H_{n'}$, where the $H_2$ space ``indicates'' the
space $H_S$ or $H_R$.

We trace out the space $H_{n'}$ in $\rho,\sigma,\rho-\sigma$. The
obtained $2\times2$ matrices are $\widetilde{\rho},
\widetilde{\sigma},\widetilde{\rho-\sigma}$. Note that the matrix
$\widetilde{\rho-\sigma}$ is diagonalized and contains the sum of all
nonnegative eigenvalues, and the sum of all negative eigenvalues on
its diagonal. Furthermore
$\widetilde{\rho-\sigma}=\widetilde{\rho}-\widetilde{\sigma}$.

Due to Lindblad-Uhlmann monotonicity of the relative von Neumann
entropy we get $S(\rho||\sigma)\ge
S(\widetilde{\rho}||\widetilde{\sigma})$. We will bound the latter by
\[1/(2\ln 2) ||\widetilde{\rho}-\widetilde{\sigma}||_{\rm 1}^2
=1/(2\ln 2) ||\widetilde{\rho-\sigma}||_{\rm 1}^2,\]
and then conclude the theorem since the trace norm of
$\widetilde{\rho-\sigma}$ is the sum of absolute values of its
eigenvalues which is the sum of absolute values of eigenvalues of
$\rho-\sigma$ by construction, i.e.,
$||\widetilde{\rho}-\widetilde{\sigma}||_{\rm 1}=||\rho-\sigma||_{\rm 1}$.

So we have to prove the theorem only for $2\times 2$ density matrices.
Assume that the basis is chosen so that $\sigma$ is diagonal. Then
\[\rho=\left(\begin{array}{ll} a&b\\b^*&1-a\end{array}\right) \mbox{ and }
\sigma=\left(\begin{array}{ll}
    c&0\\0&1-c\end{array}\right).\]

The relative von Neumann entropy
$S(\rho||\sigma)=-S(\rho)-trace[\rho\log\sigma]$. The second term is
$-a\log c-(1-a)\log (1-c)$.

The first term is minus the entropy of the distribution induced by the
eigenvalues of $\rho$. So we compute the eigenvalues.

The eigenvalues of $\rho$ are the zeroes of its characteristic
polynomial $t^2-t+a(1-a)-bb^*$. These are $1/2\pm\sqrt{1/4-a(1-a)+bb^*}$.
Let $x=1/2+\sqrt{1/4-a(1-a)+bb^*}$. Then $S(\rho)=H(x).$

The squared norm of $\rho-\sigma$ is the squared sum of absolute values of the
eigenvalues of $\rho-\sigma$. That matrix has the characteristic
polynomial $t^2-(-a(1-a)+a(1-c)+(1-a)c-c(1-c)+bb^*)$.
Thus its eigenvalues are
$\pm\sqrt{-a(1-a)+a(1-c)+(1-a)c-c(1-c)+bb^*}$. The
squared norm as squared sum of absolute values of the eigenvalues is
\[4(a^2+c^2-2ac+bb^*).\]

First we consider the case $a=c$.
To prove this case we have to show that $H(a)-H(x)\ge
2\log(e) bb^*= 2\log(e) [(x-1/2)^2-1/4+a(1-a)]=2\log(e)[(x^2-x)-(a^2-a)]$.

Considering the function $H(y)/\log(e)+2y^2-2y$, we find that it is
nonnegative and monotone decreasing for $y$ between $1/2$ and 1. Thus the inequality holds,
when $1/2 \le a$ and $a \le x$. The first condition can be assumed w.l.o.g.,
and the second condition follows from the fact that $x\ge 1/2$ is an
eigenvalue and $a\ge 1/2$ is a diagonal element.

Now we look at the case $c\ge a$. If $c<a$, we can use the same
argument for $1-c$ and $1-a$ instead. We want to show that
\[f(c)=S(\rho||\sigma)/\log(e)-\frac{1}{2}||\rho-\sigma||^2_{\rm 1}\ge 0.\]
We know this is true for $a=c$, so we show that
increasing $c$ cannot decrease the difference. This holds since:
\begin{eqnarray*}
f'(c)&=&-a/c+(1-a)/(1-c)-2(2c-2a)\\
&=& \frac{(1-a)c-a(1-c)}{c(1-c)}-4(c-a)\\
&\ge& 4(c-a)-4(c-a)\ge 0.
\end{eqnarray*}

This yields the theorem for the $2\times 2$ case and
thus in general by the previous considerations.\qed

Note that for a bipartite state $\rho_{AB}$ the following holds:
\[S(A:B)=S(\rho_{AB}||\rho_A\otimes \rho_B)\ge\frac{1}{2\ln2}||\rho_{AB}-\rho_A\otimes\rho_B||^2_{\rm 1}.\]
Thus the measurable distance between the tensor product state and the
``real'' bipartite state can be bounded in terms of the information.
We will call the value $D(A:B)=||\rho_{AB}-\rho_A\otimes\rho_B||_{\rm 1}$
the {\it informational distance}.

The next lemma collects a few properties of informational distance
that follow easily from the previous discussion.

\begin{lemm}
  For all states $\rho_{ABC}$ the following holds:
\begin{enumerate}
\item $D(A:B)=D(B:A)$.
\item $D(AB:C)\ge D(A:C)$.
\item $0\le D(A:B)\le 2$.
\item $D(A:B)\ge ||F(\rho_{AB})-F (\rho_A\otimes \rho_B)||_{\rm 1}$ for all
  completely positive and trace-preserving superoperators $F$.
\item $D(A:B)\le \sqrt{2S(A:B)}$.
\end {enumerate}
\end{lemm}

Note that lemma 2.5 implies one of the main ingredients of the round
hierarchy discovered in \cite{NTZ00} (the ``average encoding
theorem'').

Consider some density matrix $\rho_{AB}$ that is block diagonal (with
classical $\rho_A$) in the basis composed as the tensor product of the
standard basis for $A$ and some other basis for $B$. Then denote
$\rho_B^{(a)}$ the density matrix obtained by fixing $A$ to some
classical value $a$ and normalizing.  $Pr(a)$ is the probability of
$a$.

The next properties of informational distance will be used later.
\begin{lemm}
\begin{enumerate}
\item Let $\rho_{AB}$ be the density matrix of a state, where $\rho_B$
  corresponds to the density function of a classical random variable
  $B$ on $|0\rangle$ and $|1\rangle$ with $Pr(B=1)=Pr(B=0)=1/2$. Let
  there be a measurement acting on the $A$ system only and yielding a
  Boolean random variable $X$ with $Pr(X=B)\ge 1-\epsilon$ and
  $Pr(X\neq B)\le\epsilon$ (while the same measurement applied to
  $\rho_A\otimes \rho_B$ yields a distribution with $Pr(X=B)=Pr(X\neq
  B)=1/2$). Then $D(A:B)\ge 1-2\epsilon$.
\item For all block diagonal $\rho_{AB}$, where $\rho_A$ corresponds
  to a classical distribution $p_A$ on the standard basis vectors for
  $A$, the following holds:
\[D(A:B)=E_a|| \rho_B^{(a)}-\rho_B||_{\rm 1}.\]
\end{enumerate}
\end{lemm}

{\sc Proof:} For the first item observe that $D(A:B)\ge D(X:B)\ge
1-2\epsilon$.

The second item is a consequence of $D(A:B)=||\rho_{AB}-p_A\otimes
\rho_B||_{\rm 1}.$ \qed

\section{Rounds in Quantum Communication}

It is well known that for deterministic, probabilistic, (and even
limited nondeterministic) communication complexity there are functions
which can be computed much more efficiently in $k$ rounds than in
$k-1$ rounds (see \cite{DGS87}, \cite{HR93}, \cite{NW93},
\cite{PRV99}, \cite{Kl98}). In most of these results the pointer
jumping function is considered.

\begin{defi}
  Let $V_A$ and $V_B$ be disjoint sets of $n$ vertices each.
  
  Let $F_A=\{f_A|f_A:V_A\to V_B\}$, and $F_B=\{f_B|f_B:V_B\to V_A\}$.
  
  $f(v)=f_{f_A,f_B}(v)=\left\{
\begin{array}{ll}
f_A(v) & \mbox{ if } v\in V_A,\\
f_B(v) & \mbox{ if } v\in V_B.\end{array}\right.$

Define $f^{(0)}(v)=v$ and $f^{(k)}(v)=f(f^{(k-1)}(v))$.

Then $g_k: F_A\times F_B\to (V_A\cup V_B)$ is defined by
$g_k(f_A,f_B)=f_{f_A,f_B}^{(k+1)}(v_1)$, where $v_1\in V_A$ is fix.
The function $f_k: F_A\times F_B\to\{0,1\}$ is the XOR of all bits in
the binary code of the output of $g_k$.
\end{defi}

Nisan and Wigderson proved in \cite{NW93} that $f_k$ has a randomized
$k$ round communication complexity of $\Omega(n/k^2-k\log n)$ if B
starts communicating and a deterministic $k$ round communication
complexity of $k\log n$ if Alice starts. The lower bound can also be
improved to $\Omega(n/k+k)$, see \cite{Kl00}. Nisan and Wigderson also
describe a randomized protocol computing $g_k$ with communication
$O((n/k)\log n+k\log n)$ in the situation, where Bob starts and $k$
rounds are allowed. Ponzio et.~al.~show that the deterministic
communication complexity of $f_k$ is $O(n)$ then, if $k=O(1)$
\cite{PRV99}.

With techniques similar to the ones in this paper we can also show a
lower bound of $\frac{(1-2\epsilon)^2n}{2k^2}-k\log n$ for the
randomized $k$ round complexity of $f_k$ when
B starts, which is better than the above lower bounds for small
constant values of $k$.

Interaction in quantum communication complexity has also been investigated
by Nayak, Ta-Shma, and Zuckerman \cite{NTZ00}. For the
pointer jumping function their results imply the following:

\begin{fact}
$Q^{B,k}(f_k)=\Omega(n^{1/k}/k^4)$.
\end{fact}

First we give a new upper bound.  The next result combines ideas from
\cite{NW93} and \cite{PRV99}.

\begin{theor} $R_\epsilon^{k,B}(g_k)\le
  O(\frac{n}{k\epsilon}\cdot(\log^{(k/2)}n+ \log k)+k\log n)$.
\end{theor}

{\sc Proof:} First Bob guesses with public random bits
$(4/\epsilon)\cdot(n/k)$ vertices. For each chosen vertex $v$ Bob
communicates the first $\log^{(k/2)}n+3\log k$ bits of $f_B(v)$.

In round $t$ the active player communicates the pointer value
$v_t=f^{(t-1)}(v_1)$. If it's Alice's turn, then she checks, whether
$v_t$ is in Bob's list of the first round. Then Alice knows
$\log^{(k/2)}n+3\log k$ bits of $f_B(v_t)$. Note that this happens
with probability $1-\epsilon$ during the first $k/2$ rounds. In the
following assume that this happened in round $i\le k/2$, otherwise the
protocol errs.

Beginning from the round $i$, when Alice gets to know the
$\log^{(k/2)}n+3\log k$ bits of $f(v_i)$ the players communicate in
round $i+t$ for all possible values of $f(v_{i+t})$ the most
significant $\log^{(k/2-t)}n+3\log k$ bits. Since there are at most
$n/(\log^{(k/2-t)}n\cdot k^3)$ such values $O(n/k^2)$ bits
communication suffices. In the last round $v_{k+2}$ is found. Overall
the communication is at most
\[k\log n+O((1/\epsilon)\cdot(n/k) (\log^{(k/2)}n+3\log k))+k\cdot O(n/k^2).\hspace{2.2cm}\mbox{\qed}
\]

\begin{coro} If $k\ge 2\log^*(n)$ then $R^{k,B}(g_k)\le
  O((\frac{n}{k}+k) \log k)$.
\end{coro}

We can replace $k\log n$ by $k\log k$ in the above expression, because
that term dominates only if $\log k=\Theta(\log n)$.

Next we are going to prove a lower bound on the quantum communication
complexity of the pointer jumping function $f_k$, for the situation that $k$
rounds are allowed and Bob sends the first message.
We will consider a quantity $d_t$ capturing the
information the active player has in round $t$ on vertex $t+1$ of the
path. This quantity will be the informational distance between the active player's qubits
and vertex $t+1$. Our goal will be to bound $d_t$ in terms of $d_{t-1}$
plus a term related to the average information on pointers
in the other player's input. This leads to a recursion imposing a
lower bound on the communication complexity, since in the end the
protocol must have reasonably large information to produce the output,
and in the beginning the respective information is 0.

The informational distance $d_{t}$ measures the distance
between the state of, say, Alice's qubits together with the vertex $t+1$
of the path, and the tensor product of the states of Alice's qubits
and vertex $t+1$. In the
product state Alice has no information on the vertex, so if the two states are
close, Alice's powers to say something about the vertex are very limited.
We will use the triangle inequality to bound $d_t$ by the sum
of three intermediate distances.
In the first step we move from the state given by the protocol to a
state in which the $t$-th vertex
is replaced by a uniformly random vertex, independent of previous
communications. The distance to such a state
can be bounded in terms of $d_{t-1}$, because that quantity puts a bound
on Bob's ability to detect such a replacement. We use the local
transition theorem from \cite{NTZ00} to conceal Alice's ability to detect such a replacement.
Once the $t$-th vertex is random, we can
move in the next step to a state in which also vertex $t+1$ is random.
The cost of this step corresponds to the average information a player
has on a random pointer in the other player's input. The last step is
similar to the first and reverses the first one's effect, i.e.,
replaces the ``randomized'' $t$-th vertex by its real value again.We
arrive at the desired product state.

\begin{theor}
$Q^{k,B,pub}(f_k)\ge n/2^{2^{O(k)}}-k\log n$.
\end{theor}

{\sc Proof:}
Fix a quantum protocol, of the following form.
The protocol computes $f_k$ with error ${1 \over 3}$, $k$ rounds, Bob starting.
At any time in the protocol Alice has access to qubits containing her
input, some ``work'' qubits and some of the qubits used in messages so
far, the same holds for Bob. We assume that the players never change
their inputs.
Usually a protocol gets some classical $f_A$ and $f_B$ as inputs, but
we will investigate what happens if the protocol is started on a
superposition over all inputs, in which all inputs have the same
amplitude, i.e., on
\[\sum_{f_A\in \F_A,f_B\in \F_B}\frac{1}{n^n}|f_A\rangle|f_B\rangle.\]
Note that $|\F_A|=|\F_B|=n^n.$ The superposition over all inputs is
measured after the protocol has finished, so that a uniformly
random input and the result of the protocol on that input are produced.

The density matrix of the global state of the protocol is
$\rho_{M_{A,t}M_{B,t}F_AF_B}$. Here $F_A,F_B$ are the qubits holding
the inputs of Alice and Bob and $M_{A,t}$ resp.~$M_{B,t}$ are the
other qubits in the possession of Alice and Bob before the
communication of round $t$. The state of the latter two systems of
qubits may be entangled. In the beginning these qubits are independent
of the input.

We also require that in round $t$ the vertex
$v_t=f^{(t-1)}(v_1)$ is communicated by a
classical message and stored by the receiving player. 
This increases the communication by an additive $k\log n$ term. 
We demand that before round $t$ the $t$'th vertex of the path is measured
(remember that we are in a super-position over $f_A,f_B$). 
This vertex is stored in some qubits $V_{t}$.
$V_1$ has the fixed value $v_1$. 
In general, before the beginning of round $t$, we have a mixture and 
in each pure state in the
mixture the first $t-1$ vertices are fixed and $V_{t}$ is either
$F_A(v_{t-1})$ or $F_B(v_{t-1})$. We then measure $V_{t}$ in the standard
basis. The measurements do not affect the correctness of the protocol.

Let us denote $d_t=D(M_{B,t}F_B:F_A(V_t))$ when $t$ is odd,
and $d_t = D(M_{A,t}F_A:F_B(V_t))$ when $t$  is even.
In this definition, we assume that the register~$F_A(V_t)$
(or~$F_B(V_t)$) has been measured, although this measurement is not part
of the protocol.
Note that for~$t > 1$, $V_t$ is uniformly random, so (for odd~$t$) the
distance~$d_t$ is taken as the average over~$v$, of the
informational distance between the state of~$M_{B,t}F_B$ restricted
to~$V_t$ being equal to~$v$, and~$F_A(v)$, and similarly for even~$t$.

We assume that the communication complexity of the protocol is 
$\delta n$ and prove a lower bound $\delta \ge 2^{-2^{O(k)}}$.
The general strategy of the proof is induction over the rounds, to
successively bound~$d_1, d_2, \ldots, d_{k+1}$.

Bob sends the first message. As Bob has seen no message yet,
we have that $I(M_{B,1}F_B:F_A(V_1))=0$, and hence $d_1=0$. 
We show that
\begin{lemm}
\label{lem:induction}
$d_{t+1} \le 4\sqrt{d_t}+\sqrt{4\delta}$.
\end{lemm}

We see that
$d_{t+1} \le 3^t \delta^{1/2^t}$ for all $t \ge 0$.
After round $k$ one player, say Alice, announces the result
which is supposed to be the parity of $F_B(V_{k+1})$ and included in
$M_{A,k+1}$.  
On the one hand $d_{k+1} ~=~ D(M_{A,k+1}:F_B(V_{k+1}))
\le 3^{k} \delta^{1/2^{k}}$.  
On the other hand, by Lemma 3.1
$D(M_{A,k+1}:\bigoplus F_B(V_{k+1}))\ge 1-{2 \over 3}={1 \over 3}$.
Together,  ${1 \over 3} \le 3^k\delta^{1/2^{k}}$, so
$\delta \ge 2^{-2^{O(k)}}$.

We now turn to proving Lemma \ref{lem:induction}.

W.l.o.g.\ let Alice be active in round $t+1$. 
Let $M_A=M_{A,t+1}$ and $M_B=M_{B,t+1}$. 
Before the $t+1$ round $V_{t+1}=F_A(V_{t})$ is measured.
The resulting state is a
probabilistic ensemble over the possibilities to fix
$V_1,\ldots,V_{t+1}$, which are then classically distributed. 
Alice's reduced state is block
diagonal with respect to the possible values of the
vertices $V_1,\ldots,V_{t+1}$.
For any value $v$ of $V_{t+1}$ let
$\rho^{v}_{M_A M_B F_A F_B} = \rho^{V_{t+1} = v}_{M_A M_B F_A F_B}$
denote the pure state with vertex $V_{t+1}$ fixed to $v$.
We are interested in the value
\begin{eqnarray*}
d_{t+1} ~~=~~ D(M_AF_A:F_B(V_{t+1}))
 &=& \E_{v} \trnorm{\rho^{v}_{M_AF_AF_B(v)}
              - \rho^{v}_{M_AF_A}\otimes\rho_{F_B(v)}},
\end{eqnarray*}
where the distribution on vertices~$v$ (induced by the
state of the system~$F_A, F_B$) is uniform.
Recall that in the definition of~$d_{t+1}$,
$F_B(v)$ is assumed to be uniformly random (i.e., measured).
The above quantity measures how much Alice knows about the
value $F_B$ gives to the current vertex $V_{t+1}$.

We define 
\begin{eqnarray}
\label{eqn:gamma}
\gamma_{v} & \eqdef & \trn{\rho^{v}_{M_B F_B}-\rho_{M_B F_B}}.
\end{eqnarray}
I.e., $\gamma_{v}$ is the distance between
the state of Bob ($\rho^{v}_{M_B F_B}$)
before he receives the message in round $t+1$,
and the state~$\rho_{M_B F_B}$, which is his state
averaged over $v=F_A(V_t)$.
We show below that these two are almost always close to each other
(this reflects the fact that Bob does not know much about $F_A$).

For the purposes of the proof, we also
consider a run of the protocol on the uniform superposition over inputs,
where the qubits~$V_1,V_2,\ldots$ are {\em not\/} measured during the
course of the protocol. Let~$\tilde{\rho}_{M_A M_B F_A F_B}$ be 
the state before the communication in round~$t+1$ in this run of the
protocol. For any~$v \in V_B$, we define:
\begin{eqnarray}
\label{eqn:beta}
\beta_{v} & \eqdef & \trn{\tilde{\rho}_{M_AF_AF_B(v)}
                           - \tilde{\rho}_{M_AF_A}
                             \otimes \rho_{F_B(v)}},
\end{eqnarray}
where $F_B(v)$ is assumed to have been measured. Note
that~$\rho_{F_B(v)} = \tilde{\rho}_{F_B(v)}$ with this measurement; both
are randomly distributed over~$V_A$.

Let~$\rho_{M_AM_B{F_A}F_BR}$
(respectively,~$\rho^{v}_{M_AM_B{F_A}F_BR}$) be a purification
of~$\rho_{M_AM_B{F_A}F_B}$ (resp.~$\rho^{v}_{M_AM_B{F_A}F_B}$),
where $R$ is some additional space used to purify the random
path~$V_1,\ldots,V_{t+1}$ (resp.~$V_1,\ldots,V_{t}$).

We employ the following fact from \cite{NTZ00} (the ``local transition
theorem''). The fact is a variation of the impossibility result for
unconditionally secure quantum bit commitment due to Mayers \cite{M97}
and Lo and Chau \cite{LC98}.

\begin{fact}
  Let $\rho_1,\rho_2$ be two density matrices with support in a
  Hilbert space $H$, $K$ a Hilbert space of dimension at least $dim
  H$, and $|\phi_1\rangle, |\phi_2\rangle, $ any purifications of
  $\rho_1$ resp.~$\rho_2$ in $H\otimes K$. Then there is a
  purification $|\phi'_2\rangle$ of $\rho_2$ in $H\otimes K$, that is
  obtained by applying a unitary transformation $I\otimes U$ to
  $|\phi_2\rangle$, where $U$ is acting on $K$ and $I$ is the identity
  operator on $H$.
  
  $|\phi_2'\rangle$ has the property
\[||\,|\phi_1\rangle\langle\phi_1|-|\phi'_2\rangle\langle\phi'_2|\,||_1\le 2\sqrt{||\rho_1-\rho_2||_1}.\]
\end{fact}

Now, due the above fact there is a local unitary
transformation $U_v$ acting only on ${F_AM_AR}$ such
that
$$
\sigma^{v}_{M_AM_B{F_A}F_BR}
    ~~\eqdef~~ U_v \rho_{M_AM_B{F_A}F_BR} U_v^\dagger,
$$
and~$\rho^{v}_{M_AM_B{F_A}F_BR}$ are close to each other.
Moreover,
\begin{lemm}
\label{lem:rho-sigma}
For all vertices~$v \in V_B$,
\begin{eqnarray}
\label{eqn:rho-sigma}
\trn{\rho^{v}_{M_A{F_A}}  - \sigma^{v}_{M_A{F_A}}} 
    ~~\le~~ \trn{\rho^{v}_{M_A{F_A}F_B(v)}  - \sigma^{v}_{M_A{F_A}F_B(v)}}
    & \le & 2\sqrt{\gamma_{v}},\\
\label{eqn:sigma-tensor}
\trn{\sigma^{v}_{M_A F_A F_B(v)}
         - \sigma^{v}_{M_A F_A} \otimes \rho_{F_B(v)}}
    & \le & \beta_{v}.
\end{eqnarray}
\end{lemm}

We will also prove:
\begin{lemm}
\label{lem:bg}
For the uniform distribution on vertices~$v$ (induced by the state
of the system~$F_A,  F_B$),
\begin{eqnarray}
\label{lem:gamma}
\E_{v} \gamma_{v} & \le & d_t, \textrm{~~and} \\
\label{lem:beta}
\E_{v} \beta_{v} & \le  & \sqrt{4\delta}.
\end{eqnarray}
\end{lemm}

Thus, for all~$v$:
\begin{eqnarray*}
\lefteqn{ \trn{ \rho_{M_AF_AF_B(v)}^{v}
          - \rho_{M_AF_A}^{v} \otimes \rho_{F_B(v)} } } \\ [5pt]
& &
\begin{array}{cll}
    \le&  \trn{ \rho_{M_AF_AF_B(v)}^{v}
                 - \sigma^{v}_{M_AF_AF_B(v)}}&\\ [10pt]
       &   + \trn { \sigma^{v}_{M_AF_AF_B(v)}
                    - \sigma^{v}_{M_AF_A} \otimes \rho_{F_B(v)} }
          & \\ [10pt]
       &  + \trn{ \sigma^{v}_{M_AF_A} \otimes \rho_{F_B(v)}
                   - \rho_{M_AF_A}^{v} \otimes \rho_{F_B(v)} }
          &  \\ [10pt]
    \le&  4 \sqrt{\gamma_{v} }
           + \trn{ \sigma^{v}_{M_AF_AF_B(v)}
                   - \sigma^{v}_{M_AF_A} \otimes \rho_{F_B(v)}} &
           \textrm{From equation~(\ref{eqn:rho-sigma})} \\ [10pt]
    \le&  4\sqrt{\gamma_{v }}+\beta_{v} &
           \textrm{From equation~(\ref{eqn:sigma-tensor})}. 
\end{array}
\end{eqnarray*}
Finally,
\begin{eqnarray*}
\lefteqn{ D(M_AF_A:F_B(V_{t+1})) } \\ [5pt]
& &
\begin{array}{cll}
    = &   \E_{v} \trnorm{ \rho_{M_AF_AF_B(v)}^{v}
                            - \rho_{M_AF_A}^{v}
                              \otimes \rho_{F_B(v) } } & \\ [10pt]
  \le & \E_{v }[4\sqrt{\gamma_{v }}+\beta_{v}] & \\ [10pt]
  \le & 4\sqrt{\E_{v }\gamma_{v }}+\E_{v} \beta_{v} &
        \textrm{By Jensen's inequality} \\ [10pt]
  \le & 4\sqrt{d_t}+\sqrt{4 \delta}
        & \textrm{By Lemma~\ref{lem:bg}}.
\end{array}
\end{eqnarray*}
This completes the proof of Lemma~\ref{lem:induction}.

We finish the proof of Theorem 3
by proving Lemmas \ref{lem:rho-sigma} and \ref{lem:bg}.

{\sc Proof of Lemma \ref{lem:rho-sigma}:}

For equation (\ref{eqn:rho-sigma}) notice that 
\begin{eqnarray*}
\trn{\rho^{v}_{M_A{F_A}} - \sigma^{v}_{M_A{F_A}}}
&\le& \trn{\rho^{v}_{M_A{F_A}F_B(v)} - \sigma^{v}_{M_A{F_A}F_B(v)}} \\
&\le& \trn{\rho^{v}_{M_A{F_A}M_BF_BR} - \sigma^{v}_{M_A{F_A}M_BF_BR}},
\end{eqnarray*}
and by fact 6 this is at most $2\sqrt{\gamma_{v}}$.

For equation (\ref{eqn:sigma-tensor}),
\begin{eqnarray*}
\lefteqn{ \trn{ \sigma^{v}_{M_A{F_A}F_B(v)}
                - \sigma^{v}_{M_A{F_A}} \otimes \rho_{F_B(v)}} }
\\ [5pt] & &
\begin{array}{cll}
  \le & \trn{ \sigma^{v}_{M_A{F_A}RF_B(v)}
              - \sigma^{v}_{M_A{F_A}R} \otimes \rho_{F_B(v)}}
        & \\ [10pt]
  =   & \trn{ \rho_{M_A{F_A}RF_B(v)}
              - \rho_{M_A{F_A}R} \otimes \rho_{F_B(v)} }
        & \textrm{~~By unitarity} \\ [10pt]
  =   & \trn{ \tilde{\rho}_{M_A{F_A}F_B(v)}
              - \tilde{\rho}_{M_A{F_A}} \otimes \rho_{F_B(v)} }
        & ~~(*) \\ [10pt]
  =   & \beta_{v} & \textrm{~~By definition (\ref{eqn:beta})}.
\end{array}
\end{eqnarray*}
For ($*$) notice that $R$ holds the path~$V_1,\ldots,V_{t+1}$,
which is determined by~$M_A F_A$.
We can apply a unitary transformation that ``erases''
this, to give us the state~$\tilde{\rho}_{M_A{F_A}}$.
The lemma is proved.

{\sc Proof of lemma \ref{lem:bg}:}

For equation~(\ref{lem:gamma}), we have
\begin{eqnarray*}
\E_{v} \gamma_{v}
    & = & \E_{u} \trnorm{ \rho^{V_t=u}_{M_BF_B F_A(u)}
                               - \rho^{V_t=u}_{M_BF_B}
                                 \otimes \rho_{F_A(u)} } \\
    & = & D(M_B F_B : F_A(V_t)) \\
    & \le & D(M_{B,t} F_B : F_A(V_t)) ~~=~~ d_t.
\end{eqnarray*}
The last step follows from the fact that Bob sends the $t$'th message,
and this only decreases the informational distance: $D(M_{B,t+1} F_B :
F_A(V_t)) \le D(M_{B,t} F_B : F_A(V_t))$.

To derive equation~(\ref{lem:beta}),
we first bound the information Alice has on Bob's input.
\begin{lemm}
For all $t$, $I(M_{A,t}F_A:F_B) \le 2\delta n$, irrespective of whether
some registers have been measured or not.
\end{lemm}

In the beginning Alice has no information about~$F_B$,
i.e., $I(M_{A,1}F_A:F_B)=0$.
Recall that at most $\delta n$ qubits are communicated in the
protocol. Any qubit sent from Alice to Bob does not increase her
information on Bob's input. Any local unitary transformation does not
increase her information. Now assume Bob sends a qubit $Q$. Then
$I(M_AQF_A:F_B) = I(M_AF_A:F_B)+I(Q:F_B|M_AF_A) \le I(M_AF_A:F_B)+2$
due to the Araki-Lieb inequality (fact 2)
So each qubit sent from Bob to Alice
increases her information on his input by at most 2. We thus get
$I(M_{A,t}F_A:F_B)\le2\delta n$ at all times $t$. Since
measurements only decrease mutual information, the bound also holds when
certain registers are measured and others are not, during the course of
the protocol. The lemma is proved. 

Now consider the situation that $F_B$ is distributed uniformly 
instead of being in the uniform superposition (in other words,
when~$F_B$ has been measured).
Then $\E_v I(M_A F_A:F_B(v)) \le 2\delta$ (where~$v$ is uniformly
random), using that the
$F_B(v)$ are mutually independent.
Now,
\begin{eqnarray*}
\label{eqn:average-dist}
\E_{v} \beta_{v} 
    & = &  \E_v \trn{\tilde{\rho}_{M_AF_AF_B(v)}
                     - \tilde{\rho}_{M_AF_A} \otimes \rho_{F_B(v)}} \\
    & = &  \E_v D(M_{A} F_A:F_B(v)),
\end{eqnarray*}
where~$F_A M_A M_B F_B$ are as in the protocol without measurements.
Further,
\begin{eqnarray*}
\E_v D(M_{A} F_A:F_B(v))
    &\le&  \E_v \sqrt{2\, I(M_{A}F_A:F_B(v))} \\
    &\le&  \sqrt{2\, \E_v I(M_{A}F_A:F_B(v))} \\
    &\le&  \sqrt{4\delta},
\end{eqnarray*}
by Lemma 2 and Jensen's inequality.
\qed

\section{The Disjointness Problem}

We now investigate the bounded round complexity of the disjointness
problem. Here Alice and Bob each receive the incidence vector of a
subset of a size $n$ universe. They reject iff the sets are disjoint.
It is known the $Q_\epsilon^{1}(DISJ_n)\ge(1-H(\epsilon))n$
\cite{Kl00,BW01}. Furthermore $Q(DISJ_n)=O(\sqrt{n}\log n)$ by an
application of Grover search \cite{BCW98}. In this protocol
$\Theta(\sqrt{n})$ rounds are used.  By a simple reduction (see
\cite{Kl00}) we get the following result.

\begin{theor} $Q^{k,pub}(DISJ_n)=\Omega(n^{1/k})$ for $k=O(1)$.
\end{theor}

{\sc Proof:} Suppose we are given a $k$ round quantum protocol for the
disjointness problem having error $1/3$ and using communication $c$.
W.l.o.g.~we can assume Bob starts the communication, because the
problem is symmetrical. We reduce the pointer jumping function $f_k$
to disjointness.

In a bipartite graph with $2n$ vertices and outdegree 1 there are at
most $n^k$ possible paths of length $k$ starting at vertex $v_1$. For
each such path we use an element in our universe for the disjointness
problem. Given the left resp.~right side of a specific graph Alice and
Bob construct an instance of $DISJ_{n^k}$. Alice checks for each
possible path of length $k$ from $v_1$ whether the path is consistent
with her input and whether the paths leads to a vertex $v_{k+2}$ with
odd number (if the $k+1$st vertex is on the left side). In this case
she takes the corresponding element of the universe into her subset.
Bob does the analogous with his input. Now, if the two subsets
intersect, then the element in the intersection witnesses a length
$k+1$ path leading to a vertex with odd number. If the subsets do not
intersect, then the length $k+1$ path from $v_1$ leads to a vertex
with even number.

So we obtain a $k$ round protocol for $f_k$ in which Bob starts. The
communication is $c=\Omega(n)$ for any constant $k$, the input length
for the constructed instance of disjointness is $N=n^k$ and we get
$Q^{k,pub}(DISJ_N)=\Omega(N^{1/k})$ for $k=O(1)$.  \qed

\vspace{1cm}

{\bf\Large Acknowledgement}

The author wishes to thank Ashwin Nayak and Amnon Ta-Shma for
improving the presentation of theorem 3 and for helpful comments, as
well as  Pranab Sen for pointing out a mistake in an earlier version of the proof of theorem 1.

\end{document}